# Sequentially timed Framing Imaging utilizing Multiple Optical-parametric-amplification with High Spatial Resolution and Imaging Rate

Xuanke Zeng[1,2], Shuiqin Zheng[1], Yi Cai1[1], Qinggang Lin[1], Hongyu Wang[1], Xiaowei Lu[1], Jingzhen Li[1,*], Weixin Xie[2], and Shixiang Xu[1,*]

[1] Shenzhen Key Lab of Micro-Nano Photonic Information Technology, College of Physics and Optoelectronic Engineering, Shenzhen University, Guangdong 518060, China
[2] College of Information Engineering, Shenzhen University, Shenzhen, 518060, China.

**The discovery and understanding of many ultrafast dynamical processes are often invaluable. This paper is the first report to apply optical parametric amplification to implement single-shot ultrafast imaging with high space and time resolutions. Our setup breaks the constraints usually occurring among the time resolution, frame number, frame interval and spatial resolution in existing single-shot femtosecond imagers, so its performance can be greatly improved by using laser pulses with higher power and shorter time durations. Our setup has provided the highest experimental record of the frame rate and spatial resolution with the temporal resolution of 50 fs simultaneously. It is very powerful for accurate temporal and spatial observations of the ultrafast transient events with a lifetime down to the femtosecond scale, such as atomic or molecular dynamics in photonic material, plasma physics, living cells, and neural activity.**

Ultrahigh speed imaging is a powerful tool for studying ultrafast space-time dynamics in virtually the areas including science, industry, defense, energy, and medicine [1-4]. To get the space-time information of the ultrafast process accurately and reveal its dynamic law by using of ultrahigh speed imaging, the capabilities of temporal and spatial resolutions of imaging system are two important indicators. Therefore, it is the focus of research in this field to realize imaging systems with high temporal resolution, spatial resolution and real-time function [5-7]. Different types of physical processes have different time characteristics, so it is necessary to use different methods to get the ultrafast information correspondingly. Ultrafast processes can be divided into two categories, one of which is a kind of reproducible events, and the other one is a kind of non-repetitive or random events. While the pump-probe method [8-11], have the temporal resolution of a pump pulse's pulse width, is usually adopted for the former one. For the latter one, a single-shot measurement is required to acquire the sequentially timed image information of the events. In recent years, the rapid development of ultrashort pulse laser and the exploration of new principles and techniques have greatly promoted the development of ultrahigh speed imaging and circumvent the limitation of traditional imaging in a single-shot. Serial time-encoded amplified imaging/microscopy (STEAM) is an ultrafast continuous imaging method that runs at a frame rate of $\sim 10^8$ fps and opens the window onto nanosecond two-dimensional imaging with a shutter speed of $\sim 100$ ps [12]. Based on computational imaging, compressed ultrafast photography (CPU) can get a photography frequency of $10^{11}$ fps in a single shot [13]. It breaks the digitalization bandwidth limitation by incorporating a compressed-sensing-based algorithm into the data acquisition of a steak camera. But the spatial resolution is lower than 1 line pairs per mm.

By use of tomographic imaging methods, the frame rate can be up to $\sim 5 \times 10^{11}$ fps. While single-shot multispectral tomography (SMT) get a tomography image in a single shot [14], a frequency-domain tomography (FDT) [15] can acquire a tomographic image not only in a single shot but also ultrafast repetitive. However, the spatial resolution of FDT is limited to imaging of simple structured objects because the limited number of illumination angles. Sequentially timed all-optical mapping photography (STAMP) [16] is an ultrafast burst imaging method that with a high spatial resolution ($450 \times 450$ pixels) and a record high frame rate of $\sim 4.4 \times 10^{12}$ fps. However, due to the use of spectral coding for framing, the frame time and the exposure time and the number of frames are mutually affected. So when the frame time is up to 229 fs (corresponding to $\sim 4.4 \times 10^{12}$ fps), the exposure time is $\sim 800$ fs, which results in the overlap of information between adjacent frames. By utilizing spectral filtering in STAMP [17,18], a flexible increase in the number of frame is realized. The minimum frame interval of 477fs (corresponding to $\sim 2.1 \times 10^{12}$ fps) can be achieved with the exposure time equals to the frame interval.

In this paper we propose and demonstrate a single-shot burst imaging technique by utilizing multiple non-collinear optical-parametric-amplifiers (MOPA). We carried forward the frame rate of $1.5 \times 10^{13}$ fps in our experiment. This method circumvents two limitations in above imaging methods which have the temporal resolution in the picosecond to femtosecond field. One is that it has a high spatial resolution under a high frame rate of $1.5 \times 10^{13}$ fps. And the other is that the frame time, the exposure time, and the number of frames are independent of each other.

**Experimental setup**

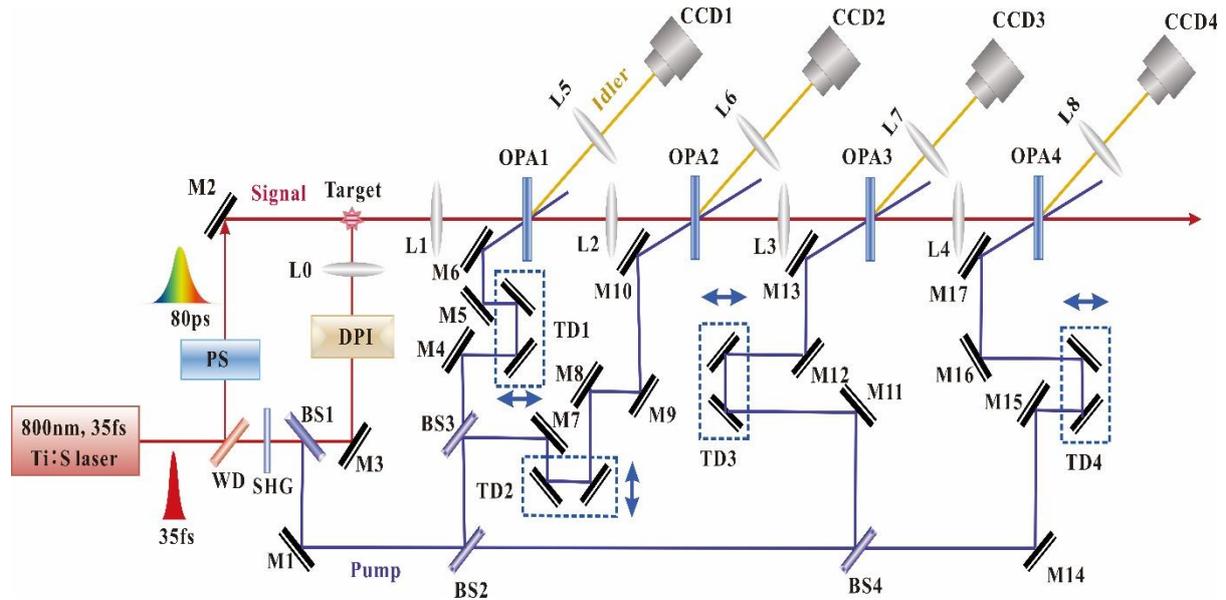

**Figure 1 | Setup for ultrafast real-time MOPA imaging.** WD: Wedge plate; PS: Pulse stretcher; SHG: Second harmonic generator; BS1~BS4: Wavelength separator; DPI: Double pulse interferometer; OPA1~OPA4: optical parametric amplifiers; M1~M17: Mirrors; MG1~MG4: temporal delay lines; L1~L4: Optical imaging lens; CCD1~CCD4: Charge-coupled device.

Figure 1 shows our experimental setup for ultrafast real-time multiple non-collinear OPA imaging. A 35 fs linear-polarized ultrashort pulse is output from a commercial Ti: sapphire laser system. The center wavelength is 800nm, and the pulse energy in single shot is 3.3mJ. 1% of the energy is separated by a wedge plate (WD) and then stretched by a pulse stretcher (PS) as a signal beam. The rest 99% of the femtosecond pulse energy is used to produce its second harmonic beam as a pump beam for OPA by a second harmonic generator (SHG). The conversion efficiency of pulse energy is about 25%. The remaining 800nm fundamental beam is separated from the second harmonic generator by a wavelength splitter (BS1), and then focus on the object plane of the OPA1 imaging system to produce an air plasma grating as a target after a double pulse interferometer (DPI). The 800nm signal beam, which is a chirp pulse with the pulse width of ~80ps after a pulse stretcher (PS), illuminates on the plasma grating. The signal beam images the target information from the object plane to the image plane with a magnified of two times by the first imaging lens L1 (a 4f system with two lens), and then reaches together with the pump beam inside the OPA1, a 29.2 °-cut 2-mm- thick type I phase matched BBO crystal, to generate an idler beam separating spatially from both of them with a small intersection angle (~2 °). The idler wavelength-converted image is recorded by a CCD camera (CCD1) with a pixel resolution of 1628×1236 pixels behind a lens (L3) which is used to image the optical information from the rear surface of the OPA1 to the CCD camera (CCD1) with two fold reduction. We enable the signal beam which carrying the transient matter information (a laser-induced plasma grating in air) and the pump beam to synchronize in time by adjusting the pump time delay (TD1) when they reach OPA1. The idler beam also carries the target information because the quantum correlation between the idler and signal [19,20]. As the idler is generated only during the 35 fs of the amplification process depending on the pump pulse width, the chirp signal beam reaching the crystal before or after the pump pulse is not amplified. Hence, a very good shutter is obtained. Similar to the OPA1 imaging, we get three other idler images by three other optical parametric amplifiers. Each OPA (BBO, the same as OPA1) is placed on the image plane successively. As a result, we get four idler images from the four OPAs by four different CCD cameras (CCD1~CCD4) respectively in a single-shot. In our setup, by adjusting each pump delay (TD1~TD4) of each OPA (OPA1~OPA4), four different time information of the air plasma grating can be recorded from the four idler beams. So the frame rate depends only on the relative time delays of the multiple OPA pump beams.

**Results**

As shown in Fig.2, the measured target is a laser-induced plasma grating in air which formed by two equal beams synchronize in time and focus with a crossing angle (~1.8 °) on the object plane of the first OPA imaging system (in Fig.1). The image is recorded by a CCD camera as shown in Fig.2a. There are four stripes in the picture and the distance between peak to peak is about 24μm, in Fig.2b, estimated by the distribution of gray value along the vertical direction of the stripes.

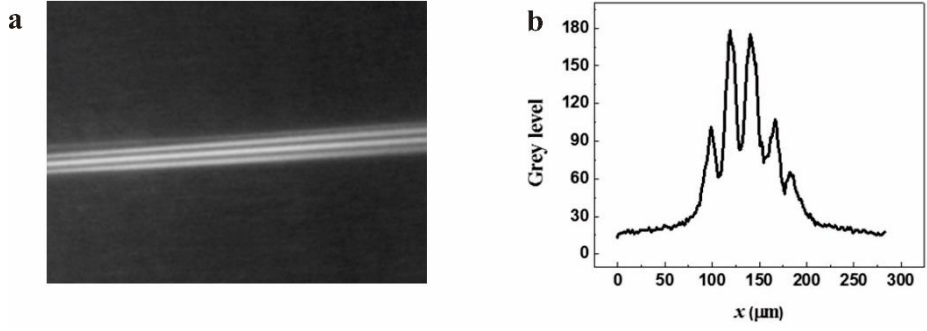

**Figure 2 | a**, The imaging target : a laser-induced plasma grating in air. **b**, the normalized gray value be perpendicular to stripes' direction.

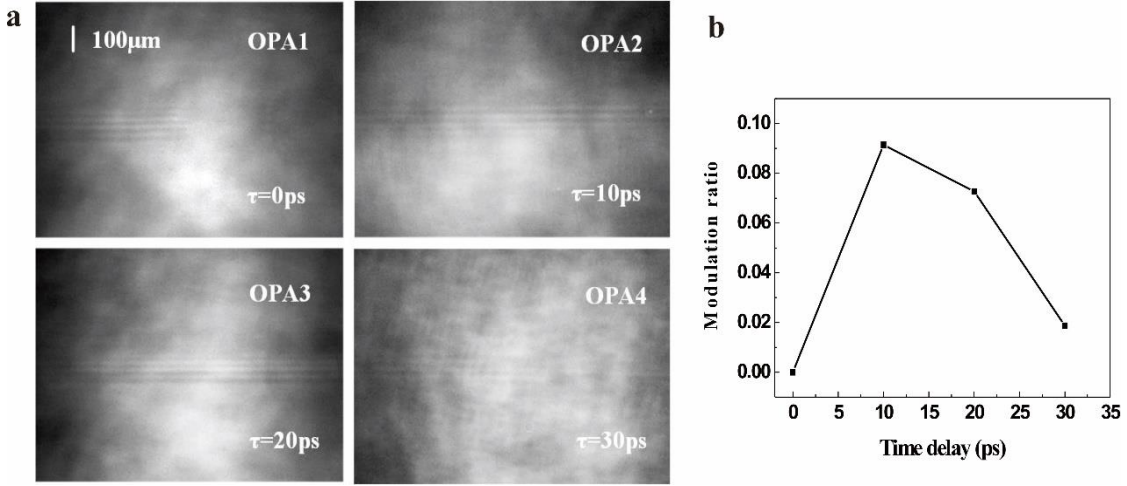

**Figure 3 | a**, Four instantaneous idler images of the plasma grating evolution at the frame rate of $10^{11}$fps (t=0, 10ps, 20ps, and 30ps). **b**, The second stripe modulation indexes versus pump time delay of the four idler images.

As depicted in Fig.1 of the setup for ultrafast real-time multiple NOPA imaging, we can get four different moments of information of the air plasma grating from four idler images by adjusting the four pump delays of OPAs. We firstly identified a 0 moment (shown in Fig. 3a of OPA1) for calibration of the four idler images, and then regulated the pump delays of the second, third and fourth OPA to 10ps, 20ps, and 30ps after the initial position in turn. In Fig. 3a, four instantaneous idler images of the plasma grating evolution at the frame rate of $10^{11}$fps (t=0, 10ps, 20ps, and 30ps) were recorded by four CCD cameras. It shows that within 0 ~ 30ps of the plasma grating excited, the stripes definition of the plasma appeared to have a decline trend after an initial ascent. That is to say, the plasma density first increased and then gradually attenuated with time in this process. To further illustrate the plasma grating image changes with the relative pump delay, we calculate the plasma grating stripe modulation index, which is defined as

$$M = \frac{I_{max} - I_{min}}{I_{max} + I_{min}}, \quad (1)$$

where $I_{max}$ and $I_{min}$ are the maximum and minimum intensity of the stripe.

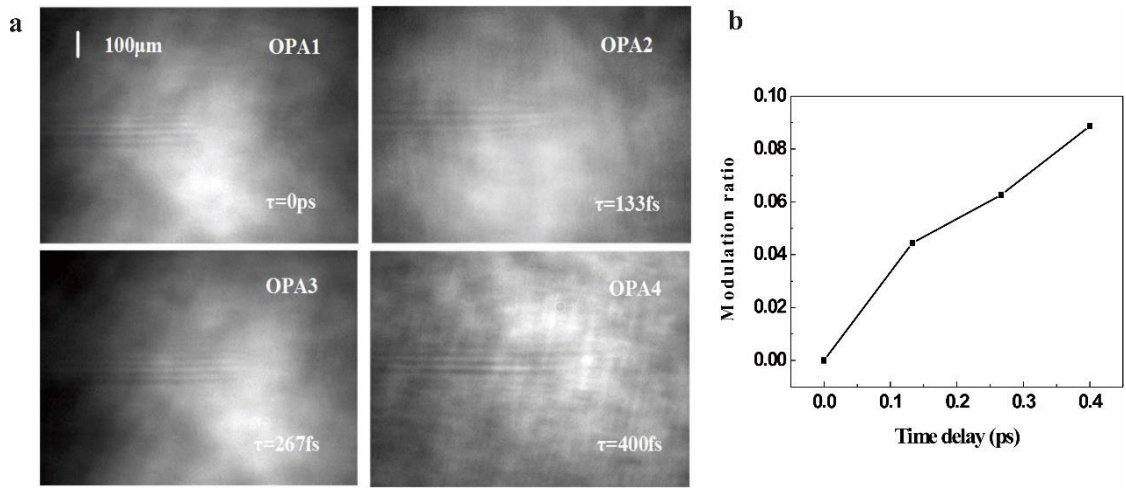

**Figure 4 | a**, Four instantaneous idler images of the plasma grating evolution at the frame rate of 7.5× $10^{12}$fps (t=0, 133fs, 267fs, and 400fs). **b**, The second stripe modulation indexes versus pump time delay of the four idler images.

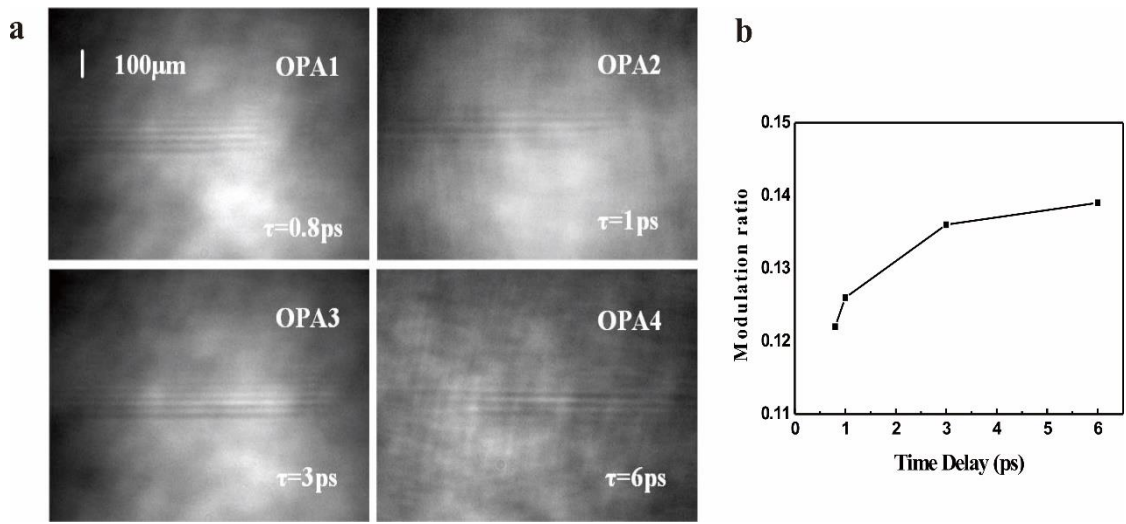

**Figure 5 | a**, Four instantaneous idler images of the plasma grating evolution with unequally frame intervals (at the moment of 0.8ps, 1ps, 3ps, and 6ps after 0 moment). **b**, The second stripe modulation indexes versus pump time delay of the four idler images.

According to equation (1), the second stripe modulation index of the target in Fig.3a changes with time is calculated and showed in Fig.3a. The stripe modulation index increased firstly and then decreased after the moment of 10ps. That is, the electron density of plasma increased firstly and then gradually attenuated with the pump delay time after the moment from 10ps to 30ps in this process. It shows that the lifetime of the excited plasma grating is about 30ps because the stripes are difficult to distinguish at the moment of 30ps. In our setup, after the 0 moment of the four OPAs is calibrated, we can get the different frame interval images just by adjusting the four pump delay in a single-shot. In order to get more accurate information of plasma grating, we regulated the pump delays of the second, third and fourth OPA to 133fs, 267fs, and 400fs after the initial position in turn. So four instantaneous idler images of the plasma grating evolution with the equally frame rate of 7.5×$10^{12}$fps were recorded by the CCD cameras as shown in Fig.4a. One can see that the stripes gradually appeared and the definition of the plasma to have a rise. So we can get that the electron density of plasma gradually increased from the moment of 0 to 400fs. The second stripe modulation index of the plasma grating, is calculated according to equation (1) and showed in Fig.4b. This shows that the electron density of the plasma grating is increasing monotonously. Since the frame intervals depends only on four independent pump delays, we can not only obtain a motion-picture with equally frame intervals, but also with unequally frame intervals according to demand. In Fig. 5a, four idler images were recorded with unequally frame intervals by adjusting the four pump time delays (0.8ps, 1ps, 3ps, 6ps). The stripes definition of plasma grating in Fig.5a is better than in Fig.4a. So we can get that the plasma density gradually increased after 400fs the plasma grating excited. The second stripe modulation index of the plasma grating, is calculated and showed in Fig.5b, also rises monotonously. That is to say, the electron density of the plasma grating is still increasing from the time of 0.8ps to 6ps.

As a result, just by adjusting the pump relative delays of the four OPAs, a motion-picture with equally or unequally frame intervals can be constructed without repeated measurement. And then, we recorded a rotating optical lattice with a rotating rate of 13.5 Trad/s, and sue our setup to get four idler images with 67fs, that corresponding highest frame rate is up to $1.5 \times 10^{13}$ fps in our experiment. The frame rate, which depends on the width of pump pulse and the accuracy of the displacement platform, can be further improved according to the needs of imaging.

**Discussion**

**High spatial resolution of ultrafast MOPA imaging.** Our setup, the ultrafast MOPA imaging technique, circumvents two limitations in former imaging methods which have the temporal resolution in the picosecond to femtosecond field. One is that it has a high spatial resolution under a frame rate as high as $1.5 \times 10^{13}$ fps. It's well known that the interacting beams in OPA imaging must be satisfy the conditions of energy conservation and momentum conservation. The spatial frequency bandwidth [21,22], which depends on the crystal thickness and pump energy, is the main factor affecting the spatial resolution in OPA imaging. Fig.6 shows the parametric gain versus spatial frequency of a 29.2°-cut type-I phase-matching $\beta$-BBO at a pump level of 30 GW/cm$^2$. The peak value of OPA gain is up to $1.1 \times 10^5$ and the gain bandwidth at 3 dB is about 28 line pairs per millimeter (lp/mm). The high gain of OPA provides enough energy for imaging to reduce the sensitivity of the CCD camera, and the spatial frequency bandwidth enables the spatial resolution of OPA imaging greater than 25 lp/mm theoretically. We measured the spatial resolution which is higher than 22 lp/mm in the vertical direction by recording the target of a USAF 1951(0) test pattern in previous experiment [23]. As a result, the spatial resolution of MOPA imaging meets the requirements for the target of a plasma grating which has a periodic structure in the vertical direction of 24μm and magnified two times in the position of the first OPA imaging plane. In addition, since each image is recorded on a separate camera, the size of the frames depends on the pixel size of the camera, so that it is easy to obtain a large space-bandwidth-product.

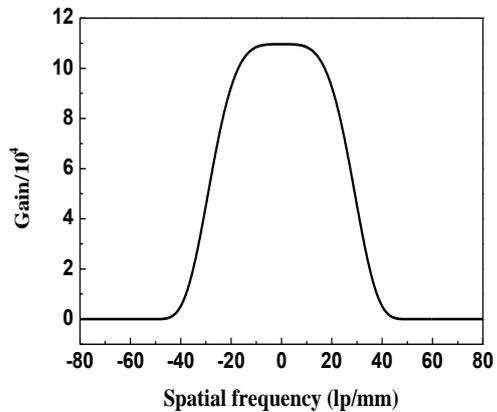

**Figure 6** | The parametric gains vs. spatial frequency of a non-collinear type-I phase-matching β-BBO OPA.

**The frame time, the exposure time, and the number of frames are independent of each other.** As we know, a high speed photography system is a complex measuring information system in time domain, and it is also a stochastic system for plenty of measuring points. In principle, the frame time, the exposure time and the number of frames should be independent of each other in a desired high speed imaging system. However, the above mentioned three parameters in some methods will depend on each other when the frame rate is as high as $10^{11}$fps ~ $10^{12}$fps. Such as the STAMP [16], achieves a record frame rate of $10^{12}$ fps by using the spectral coding method which will result in the dependence among the frame time, the exposure time, and the number of frames because of the uncertainty relation between spectrum and time. So the number of frames will be restricted in such a high frame rate. Besides, the modified temporal qualify factor $g^{2/3}$ [24,25], which is defined as the ratio between frame time and effective exposure time, is less than 1 since the exposure time is ~800fs when the frame time is up to 229 fs (corresponding to ~$4.4\times10^{12}$fps). It means that the information between the adjacent frames is overlapping and the time resolution is insufficiency at such high frame rate.

In our MOPA method, each image comes from an independent idler beam of each optical parametric amplifier. As a result, the frame time (frame rate) depends on the relative pump delays of the optical parametric amplifiers while the frame time, the exposure time and the number of frames are independent of each other. In Fig.7, the exposure time $\Delta\tau$ (i.e. the full width at half maximum of pump pulse), which is measured by a SPIDER setup, is approximately equal to 40fs. As the minimum framing time $\Delta t=133$fs (corresponding to $7.5\times10^{12}$fps) is obtained in our setup, the modification temporal factor $g^{2/3}$ can be calculated to be 2.23 which can help to avoid the overlap of information between adjacent frames. And the frame rate can be further improved by optimizing the MOPA imaging system. In addition, the number of frames can be increased freely without affecting the frame time and exposure time by designing a more compact optical system theoretically.

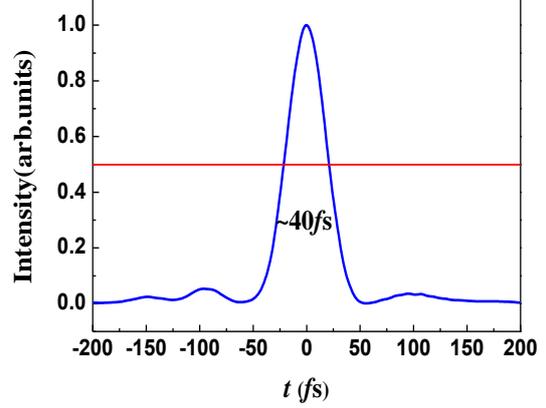

**Figure 7** | The measurement of the exposure time $\Delta\tau$ by a SPIDER setup.

In summary, this paper reports a novel setup to visualizing evolving ultrafast transient matters in a single-shot by MOPA. Instantaneous time evolution of a laser-induced plasma grating in air (~ 24 μm periodic structures) is visualizing at a maximum imaging frame rate as high as $7.5\times10^{12}$ fps with a pixel resolution of 680×680 pixels in the experiment. Due to the frame time, the exposure time, and the number of frames are independent of each other, the exposure time can be maintained and the number of frames can be increased freely at a frame rate of $7.5\times10^{12}$ fps or even higher by optimizing the imaging system. This imaging method provides a new way for blur-free observation of fast transient dynamics in virtually areas including plasma physics, hydrodynamics, photochemistry and biomedical medicine.

**Methods**
**Spatial frequency bandwidth of OPA imaging.** The gain for undepleted plane pump approximation can be written as [21]

$$G_s = |\mu_q|^2, \ G_i = |v_q|^2, \qquad (2)$$

where the subscripts '*s*' and '*i*' stand for the signal and idler, $q$ is for spatial frequency (lines/mm). The coupling coefficients $\mu$ and $\nu$ are given by

$$\mu(q) = \left[\cosh(hl) + i\frac{\Delta k_{eff}}{2}\frac{l\sinh(hl)}{hl}\right] \times \exp\left(-i\frac{\Delta k_{eff}}{2}l\right), \quad (3)$$

$$\nu(q) = \left[i\frac{\kappa l}{2}\frac{\sinh(hl)}{hl}\right] \times \exp\left(i\frac{\Delta k_{eff}}{2}l\right),$$

where $h = [\kappa^2 - (\Delta k_{eff})^2]^{1/2}/2$, $\Delta k_{eff} = k_p - k_s - k_i + 2\pi^2 q^2 (k_s^{-1} + k_i^{-1})$ is the effective phase mismatch, $\kappa$ is the parametric gain coefficient which is proportional to the intensity of the pump beam, and $l$ is the length of parametric crystal. The spatial frequency bandwidth $\Delta f$ of OPA gain at 3dB can be written as [26]

$$\Delta f = \frac{1}{\pi}\left(\frac{k_1^2 \kappa \ln 2}{4l}\right)^{\frac{1}{4}}. \quad (4)$$

The equation (2) ~ (4) shows that thin crystal thickness with strong pump intensity is helpful to realize a large spatial frequency bandwidth for OPA imaging. However, excessive pump intensity will damage the crystal, while thin crystal thickness will make the gain reduction.

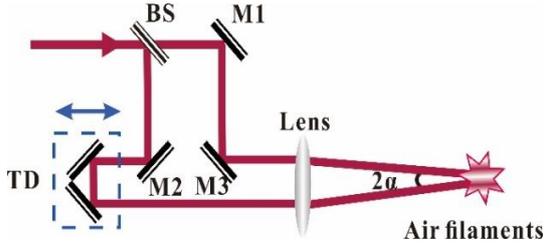

**Figure 8 | Diagram of DPI for generating ultrafast events.** BS: 50/50 beam splitter, M1~M3: mirror, TD: Time delay.

**Generation of plasma grating.** Fig.8 is the diagram of DPI in Fig.1 for generating ultrafast events. The two equal beams are split from the 800nm femtosecond pulse beam by a 50:50 beam splitter (BS). The reflected beam, which passed through a high reflectivity mirror (M2) and a time delay (TD), incident to an optical lens together with the transmitted beam which passed though the high reflectively mirrors (M1 and M3) in a small angle to induce air plasma. When the two beams are synchronized in time by adjusting the optical time delay (TD), the plasma grating of a periodic structure is formed due to interference. The spacing of the interference fringe can be written

$$\Delta = \frac{\lambda}{2\sin\alpha}, \quad (5)$$

where $\lambda$ is the wavelength of the femtosecond laser pulse, $2\alpha$ is the incident angle of the two beams.

**Acknowledgments**

This work was supported by National Natural Science Fund of China (61775142, 61827815, and 61705132), China Postdoctoral Science Foundation (2017M612726), Shenzhen basic research project on subject layout (JCYJ20170412105812811), and the Fund of the International Collaborative Laboratory of 2D Materials for Optoelectronics Science and Technology, Shenzhen University (2DMOST2018019).